# In plane reorientation induced single laser pulse magnetization reversal in rare-earth based multilayer

Y. Peng[1], D. Salomoni[2], G. Malinowski[1,*], W. Zhang[1,3], J. Hohlfeld[1], L. D. Buda-Prejbeanu[2], J. Gorchon[1], M. Vergès[1], J. X. Lin[1], R.C. Sousa[2], I. L. Prejbeanu[2], S. Mangin[1] and M. Hehn[1,*]

[1] *Université de Lorraine, CNRS, IJL, F-54000 Nancy, France*
[2] *Spintec, Université Grenoble Alpes, CNRS, CEA, Grenoble INP, IRIG-SPINTEC, 38000 Grenoble, France*
[3] *Anhui High Reliability Chips Engineering Laboratory, Hefei Innovation Research Institute, Beihang University, Hefei 230013, China. MIIT Key Laboratory of Spintronics, School of Integrated Circuit Science and Engineering, Beihang University, Beijing 100191, China*

\* *michel.hehn @univ-lorraine.fr, gregory.malinowski@univ-lorraine.fr*

**Abstract**

Single Pulse All Optical Helicity Independent Switching (AO-HIS) represents the ability to reverse the magnetic moment of a nanostructure using a femtosecond single laser pulse. It is an ultrafast method to manipulate magnetization without the use of any applied field. Since the first switching experiments carried on GdFeCo ferrimagnetic systems, single pulse AO-HIS has been restricted for a while to Gd-based alloys or Gd/FM bilayers where FM is a ferromagnetic layer. Only recently has AO-HIS been extended to a few other materials: MnRuGa ferrimagnetic Heusler alloys and Tb/Co multilayers with a very specific range of thickness and composition. Here, we demonstrate that single pulse AO-HIS observed in Tb/Co results from a different mechanism than the one for Gd based samples and that it can be obtained for a large range of rare earth–transition metal (RE–TM) multilayers, making this phenomenon much more general. Surprisingly, in this large family of (RE–TM) multilayer systems, the threshold fluence for switching is observed to be independent of the pulse duration, up to at least 12 ps. Moreover, at high laser intensities, concentric ring domain structures are induced, unveiling multiple fluence thresholds. These striking switching features, which are in contrast to those of AO-HIS in GdFeCo alloys, concomitant with the demonstration of an in-plane reorientation of the magnetization, point towards an intrinsic precessional reversal mechanism. Our results allow expanding the variety of materials with tunable magnetic properties that can be integrated in complex heterostructures and provide a pathway to engineer materials for future applications based on all-optical control of magnetic order.

**Introduction**

The development of new strategies to perform magnetization reversal at ultra-short timescales fits with the ceaseless demand of future magnetic storage for non-volatile, energy efficient and ultrafast functional memory or logic elements. Historically, the writing of information on magnetic media was performed using a magnetic field. However, this became limiting when reducing the bit size below the hundred nanometer scale and increasing the writing speed above the GHz. Alternative solutions have emerged using spin-polarized currents in nanosized magnetic structures like the spin transfer torque switching [KAT00] or more recently spin-orbit torque switching [MIR11]. Nevertheless, these technologies are limited due to the large increase of the required current density when decreasing the pulse duration, precluding their potential use below 100 ps timescales [LIU10, GAR14].

In 1996, a major discovery of Bigot *et al.* launched the new research area of femtomagnetism after demonstrating that a femtosecond laser pulse excitation of a thin Ni film leads to a sub-picosecond demagnetization [BIG96]. Almost ten years later, a complete deterministic all-optical switching (AOS) of the magnetization of ferrimagnetic GdFeCo alloys was demonstrated using circularly-polarized laser pulses [STA07]. Subsequently, it was shown in similar GdFeCo layers that a single femto-second laser pulse could induce a toggle switching independently of the light helicity. The effect named All Optical- Helicity Independent switching (AO-HIS) results from an ultrafast heating process related to distinct dynamics of both the rare earth and transitions metals elements and related to the high transient electron temperature being out of equilibrium with the lattice [RAD11,OST12]. Nevertheless, the possibility to reverse the magnetization in such alloys using optical or electrical excitations with pulses longer than the electron-phonon relaxation time challenges a mechanism involving the necessity of a strong nonequilibrium electronic state [STE11,GOR16,YAN17,DAV20a,WEI21]. Davies *et al.* recently pointed out the importance and the competition between the inter-sublattice exchange coupling and the spin-lattice relaxation time [DAV20a]. This could potentially explain why AO-HIS has for a long time only been observed in GdFeCo alloys [OST12] and Gd/Co multilayers [LAL17]. Recently, AO-HIS was reported in half-metallic ferrimagnetic Heusler alloys $Mn_2Ru_xGa$ which possesses two inequivalent Mn sublattices [BAN20]. The results suggested that the magnetization switching is exchange-driven in agreement with what has been previously reported for the case of GdFeCo [ATX14,DAV20a,DAV20b].

Surprisingly, Avilés-Felix and co-workers reported the possibility to reliably toggle switch the magnetization of a Tb/Co multilayer, whose anisotropy is large enough to preserve magnetically stable information bits at small diameters in a magnetic tunnel junction and maintain its perpendicular anisotropy even after a 250°C annealing process [AVI19,AVI20]. Using a crossed-wedge multilayer structure, they showed that the switching occurs in a very narrow range of Tb and Co thicknesses, that has to be precisely controlled at the Angstrom level, only when a multilayer is used and for pulse durations depending on the Tb and Co thickness ratio. These striking results potentially open the way for their integration into scalable magnetic tunnel junctions but the mechanism leading to single pulse AO-HIS in these samples remains unclear.

In this paper, we present a detailed and extensive study regarding single pulse AOS of magnetization in a large variety of samples such as magnetic multilayers and alloys containing different rare earth (RE) (Tb, Dy) and transition metals (TM) (Co, Py and Fe) elements. We demonstrate that a much more robust toggle switching is obtained when Tb is replaced by ferrimagnetic alloys with thicknesses up to few tens of nanometers and for laser pulses as long as 12 ps. **Furthermore, we show that the magnetization reversal process is completely different from the one observed in GdFeCo or Mn$_2$Ru$_x$Ga alloys and involves an in-plane magnetization reorientation and probably a precessional mechanism.**

As a starting point in our study, we consider [Tb/Co]$_5$ and [Tb/Fe]$_4$ magnetic multilayers, similar to the one used by Avilés-Felix *et al.* [AVI19,AVI20] (5 and 4 is the number of repetitions). These multilayers exhibit a strong perpendicular magnetic anisotropy at room temperature and the magnetization of the Tb and Fe or Co layers are antiferromagnetically exchange coupled at the interface. Typical results obtained for a [Tb(1.31 nm)/Fe(1.89 nm)]$_4$ multilayer are presented in Figure 1 (the study as a function of thickness of Tb and Fe is reported in Figure S1). Tb/Co multilayers present the same behavior as reported in Figure S2-a (the study as a function of thickness is shown in Figure S2-b). As depicted in Figure 1 (a), toggle switching of a single magnetic domain is observed for an excitation with a pulse duration of 50 fs and a low fluence of 1.7 mJ/cm$^2$. More interestingly, for laser fluence larger than 2.6 mJ/cm$^2$, a bullseye structure starts appearing with opposite magnetic directions in adjacent rings. Increasing the laser fluence results in a higher number of rings. As shown in Figure 1(b), the central domain is alternatively being reversed and this up to the higher fluence (7.0 mJ/cm$^2$) for which a demagnetized state is stabilized. This striking behavior has been studied systematically as a function of the laser pulse duration, which allows us to establish a complete state diagram (pulse duration versus fluence) reported in Figure 1(c). This state diagram resembles neither the state diagram observed in the case of AO-HIS in GdFeCo [WEI21] nor the state diagram observed in the case of multiple pulse All Optical Helicity Dependent Switching (AO-HDS) [KIC19]. Surprisingly, the fluences required to reverse and stabilize a given number of rings depend very little or not at all on the duration of the laser pulses. Single laser pulse induced magnetization switching is observed for pulse durations up to 12 ps, the longest pulse duration reachable in our ultrafast laser equipment. Both Tb/Co and Tb/Fe multilayers behave similarly, both having ring structures and low dependence of the reversal fluences on the laser pulse duration. Both features will be considered to determine the conditions to obtain the reversal and to discuss the possible mechanism at its origin.

The ring structure does not seem to have a dipolar origin i.e . linked to the magnetostatic coupling between adjacent magnetic domains. As reported in figure 2, we performed 2 successive pulses by slightly moving the spot: the local magnetic state is found to be independent of the configuration of the neighbouring first domains and depends only on the distance to the spot center. For the fluence used in the experiment shown in figure 2, region 2 experiences 2 reversals, one for each pulse, region 1 (respectively 1') experiences reversal for pulse 1 (respectively 2) and the state of region 0 does not change. Similar ring structures have been observed in ferrites and has been attributed to the reorientation of the magnetization from out of plane to in plane followed by its precession [DAV19, SHE18]. In metallic thin films [KAT16], magnetization precession could be observed without magnetization reversal. In both cases, an in plane magnetic field was applied which is not the case in our study.

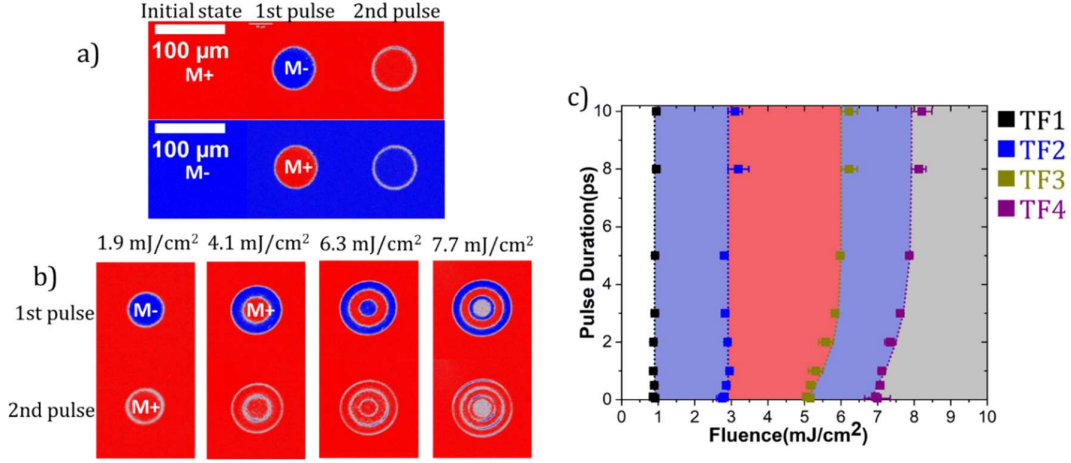

**Figure 1 :** Single pulse switching and state diagram for [Tb(1.31 nm)/Fe(1.89 nm)]$_4$ multilayer under zero applied field. M+ (respectively M-) corresponds to magnetization pointing perpendicular to film plane, along +z (resp. -z) in red (resp. in blue). a) Background subtracted images after each single pulse with 1.7 mJ/cm$^2$ laser pulse at 50fs; b) Background subtracted images after first single with 50 fs laser pulse of different fluence; c) State diagram of reversal pulse duration versus fluence. TF1 is the border for switching; TF2 is the border to two domains state (one ring); TF3 is the border to three domains state (two rings); TF4 is the border to demagnetized state at the center (three rings). The dotted lines are guide for the eyes.

The fact that the critical fluences (TF$_1$, TF$_2$,…) are independent of pulse duration, implies that the underlying mechanism is not governed by the maximum electronic temperature which decreases with increasing the pulse duration for a given fluence but by the energy transferred to the lattice. Ultimately, what matters is the total amount of energy pumped into the system, independently of its rate. Therefore, the energy density after equilibration of electrons-spins-phonons is likely to govern the reversal. To go deeper into this intuitive description, we performed two-temperature (2T) model simulations [MOR17] (supplementary material). As expected, increasing the fluence rises both electron, $T_e$, and phonon, $T_{ph}$, temperatures but, with constant fluence, the $T_e$ peak reduces drastically upon increasing the pulse duration while $T_{ph}$ remains constant. Increasing $T_{ph}$ above a threshold could lead to a modification of the magnetic anisotropy and induce the magnetization precessional switching for which a full demagnetization of the TM is not mandatory. The transfer of energy to the phonon is here justified by the strong spin orbit coupling present for Tb or Dy, contrary to Gd.

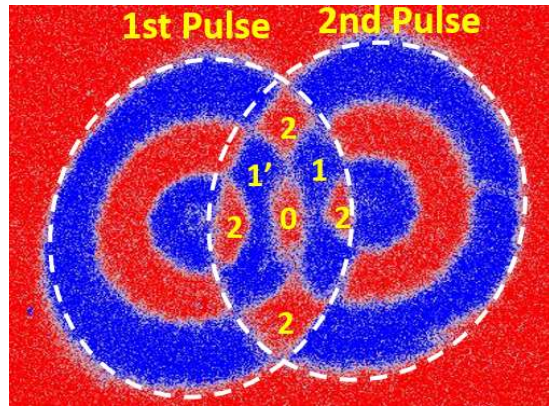

**Figure 2 :** Domain pattern obtained by sending 2 successive pulses: a first one and then, after moving the laser at a distance smaller than the spot size, a second one. The two spots are then over lapping. Experiment done on [Tb(1.31 nm)/Fe(1.89 nm)]$_4$ multilayer. The color scale corresponds to magnetization pointing out of plane: M+ (respectively M-) along the perpendicular to the film plane, along +z (resp. -z) in red (resp. in blue). The sample is initially saturated along the M+ direction. The laser pulse has a fluence of 5.2 mJ/cm$^2$ and a duration of 50fs.

In the following, we will test the hypothesis of a spin reorientation accompanied by a precession, highlighting the elements required to obtain a reversal in a single pulse and optimize the material parameters. In the case of the Tb/Co and Tb/Fe multilayers, Tb is magnetic at room temperature only because of its proximity or contact with the Co or Fe layers. Thus, modifying its thickness will also lead to a variation of its anisotropy and its magnetic moment. In order to check if thicknesses or the variation of the magnetic properties are key parameters in the process as reported in [AVI19,AVI20], the Tb layer was replaced by a rare earth–transition metal (RE–TM) alloy with a high RE concentration so the net magnetization of the alloy points into the direction of the magnetization of the RE sublattice. Due to the high RE content, the alloy has a Curie temperature higher but close to room temperature [HAN89] and possess a perpendicular to film plane magnetic anisotropy (PMA). We will compare the measurement results obtained on [Tb/Co], the multilayer in which this reversal has been discovered [AVI19,AVI20], and on [Tb/Fe] multilayer reported here to [TbCo/Co], [TbCo/Py], [TbCo/CoFeB], [DyCo/Co] [TbFe/Co] multilayers. The complete study indicates that all these combinations of multilayers are showing a similar all optical switching behavior, i.e. a single pulse magnetization reversal with ring structure and a magnetization reorientation in plane compatible with a precessional process for the reversal. The list of all the sample tested is given in table 1.

| # | Stacks | | 0.8 nm | 1.2 nm | 1.6 nm | 2.0 nm | 2.4 nm | 2.8 nm | 3.2 nm |
|---|---|---|---|---|---|---|---|---|---|
| 1 | [Tb(wedge)/Fe(wedge)]*4 | | | ■ | ■ | | | | |
| 2 | [Tb(wedge)/Co(wedge)]*5 (Cross wedge) | | | | ■ | | | | |
| 3 | [Tb$_{40}$Co$_{60}$(x)/Co(wedge)]*3 | x = 1 | | | | | ■ | | |
| 4 | | x = 2 | | | | | ■ | | |
| 5 | | x = 3 | | | | | ■ | ■ | |
| 6 | | x = 4 | | | | | ■ | ■ | |
| 7 | | x = 5 | | | | | | ■ | ■ |
| 8 | | x = 7,5 | | | ■ | ■ | ■ | ■ | |
| 9 | [Dy$_y$Co$_{100-y}$ (4) /Co(wedge)]*3 | y = 25 | | | | | | ■ | |
| 10 | | y = 30 | | | | ■ | ■ | | |
| 11 | | y = 35 | | | | ■ | | | |
| 12 | Pt(5) /Co(wedge)/ Tb$_{32}$Co$_{68}$ (4) | | | | | | | | |
| 13 | Pt(5) /Co(wedge)/ Tb$_{32}$Co$_{68}$ (8) | | | | | ■ | | | |
| 14 | Pt(5) /Co(wedge)/ Tb$_{32}$Co$_{68}$ (16) | | | | | | | | |
| 15 | Dy$_{35}$Co$_{65}$ (4) /Co(wedge)/ Dy$_{35}$Co$_{65}$ (4) | | | | | | | ■ | |
| 16 | Tb$_{32}$Co$_{68}$ (4) /Co(wedge)/ Tb$_{32}$Co$_{68}$ (4) | | | | ■ | ■ | | | |
| 17 | Tb$_{32}$Co$_{68}$ (4) /CoFeB(wedge)/ Tb$_{32}$Co$_{68}$ (4) | | | ■ | ■ | ■ | | | |
| 18 | Tb$_{55}$Fe$_{45}$ (4) /Co(wedge)/ Tb$_{55}$Fe$_{45}$ (4) | | | | | ■ | | | |
| | **Thickness of the TM layer** | | 0.8 nm | 1.2 nm | 1.6 nm | 2.0 nm | 2.4 nm | 2.8 nm | 3.2 nm |
| 19 | Tb$_{32}$Co$_{68}$ (4) /Permalloy (wedge)/ Tb$_{32}$Co$_{68}$ (4) | | | | | | ■ | ■ | ■ |
| | | | 2.0 nm | 2.6 nm | 3.2 nm | 3.8 nm | 4.4 nm | 5.0 nm | 5.6 nm |

**Table 1:** Stacks that have been studied in this paper. 1-11 are multilayers, 12-14 are bilayers and 15-19 are trilayers. Light pink color shows the TM layer (including Co, Fe, CoFeB, Permalloy) thickness range that have been studied in this paper, while blue color shows the region of TM thickness where single shot switching occurs in different stacks. Note that stack 1 is a double wedge-sample in the same direction, while stack 2 is a cross-wedge sample. Numbers in parenthesis are the layer thicknesses in nanometers.

Finally, if the in plane reorientation of the magnetization of the ferromagnetic layer (Py, CoFeB or Co) is a key ingredient to promote the toggle reversal of the magnetization of the whole multilayer stack, PMA anisotropy in the FM layer has to vanish during the process. While Co, CoFeB or Py layers in contact with a buffer or capping Pt layer or with an insulating MgO layer have high anisotropy, a multilayer structure appears necessary for the proposed switching mechanism. Since PMA in TbCo, DyCo or TbFe alloys is expected to disappear as the lattice temperature rise after the laser pulse, the simplest multilayers to switch are then TbCo/X/TbCo (X = Co, Py, CoFeB), Y/Co/Y (Y = DyCo, TbFe) tri-layers in which Co, Py, CoFeB PMA is induced by its proximity with TbCo, DyCo or TbFe alloys.

The results obtained on Tb$_{32}$Co$_{68}$/Co(1.52 nm)/Tb$_{32}$Co$_{68}$ trilayer are reported in figure 3 while those on the Tb$_{32}$Co$_{68}$/Co(wedge)/Tb$_{32}$Co$_{68}$, Dy$_{35}$Co$_{65}$/Co(wedge)/Dy$_{35}$Co$_{65}$, Tb$_{55}$Fe$_{45}$/Co(wedge)/Tb$_{55}$Fe$_{45}$, Co$_{67}$Tb$_{32}$/Py(wedge)/Co$_{67}$Tb$_{32}$ and Tb$_{32}$Co$_{68}$/CoFeB(wedge)/Tb$_{32}$Co$_{68}$ tri-layers are given in supplementary figure S4-a to S4-e respectively. In all those tri-layers, the thickness of the ferrimagnetic alloy was fixed to 4nm. In Tb$_{32}$Co$_{68}$/Co(wedge)/Tb$_{32}$Co$_{68}$, single pulse reversal can occur for Co thickness above 1.81 nm. The reversal is repeatable after 30000 laser pulses and the ring structure is present. For Co thickness less than 1.5 nm, only thermal demagnetization has been observed. The same trend is seen for TbCo, DyCo and TbFe based trilayers, where a minimum FM thickness is required to obtain the reversal (see supplementary Figure S4). Our findings agree with results reported in the literature, for TbCo or DyCo single layer alloy [HU22] which does not demonstrate single pulse AO-HIS. However, the presence of an additional exchange coupled FM layer makes single pulse AO-HIS possible. Very importantly, our findings demonstrate undoubtedly that we can largely extend the domain of thicknesses in which the single laser pulse reversal exists. In $Tb_{32}Co_{68}$ (4 nm)/Co(t nm)/ $Tb_{32}Co_{68}$ (4 nm), single pulse reversal appears for t ranging from 1.5 to 2.3 nm. With the use of CoFeB (respectively Py), this thickness can be extended up to 2.87nm (respectively 5.6 nm). This is well beyond the limited range of thicknesses for which the AOS switching was observed for Tb/Co multilayers.

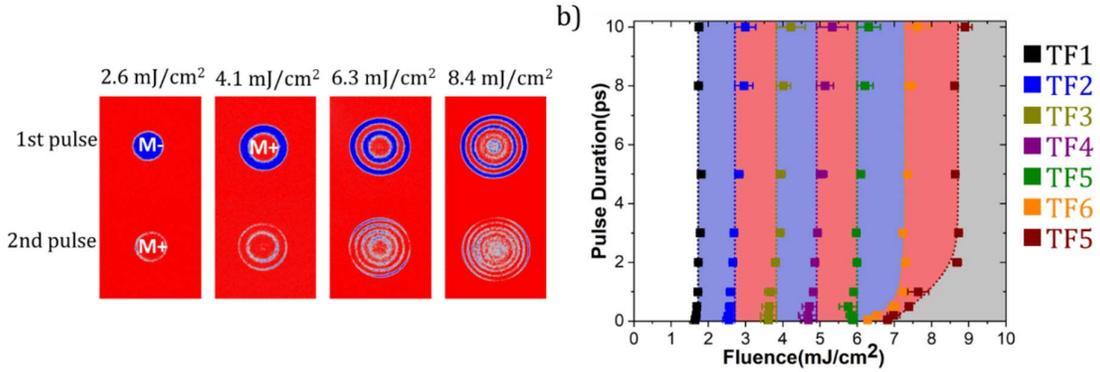

**Figure 3** : Single pulse reversal in $Tb_{32}Co_{68}(4)/Co(1.52nm)/Tb_{32}Co_{68}(4)$ trilayer. a) Background subtracted images after first and second pulses with 50 fs laser pulse with different fluences; c) State diagram which is plotted by pulse duration as a function of fluences. TF1 is the border for switching; TF2 is the border to two domains state, one ring; TF3 is the border to three domains state, two rings; etc….TF7 is the border to demagnetized state at the center. The dotted lines are guide for the eyes.

One common feature of all trilayers and multilayers, which exhibit the single laser pulse reversal, is the reorientation of the magnetization in the plane. In Figure 4, we report the magnetization dynamics for three different systems, the historical Tb/Co multilayer as well as the TbCo/Co/TbCo trilayers and [TbCo/Co] and [DyCo/Co] multilayers. This dynamic behavior is obtained by measuring the Time Resolved Magneto Optic Kerr images (TR-MOKE). The polar geometry used in this measurement allows mainly to probe the perpendicular to film plane component of transition metal magnetization. The stroboscopic measurement requires to reset the magnetic configuration after each pulse. Therefore, a DC magnetic field applied during the measurement. The measurements show that after a fast demagnetization at a time scale of some 100 fs, there is first a recovery of the magnetization, similar to the conventional demagnetization/re-magnetisation shown in various ferromagnetic and ferrimagnetic materials. At longer timescales, a few tens of picoseconds, a magnetization decrease is observed. This decrease cannot be attributed to the creation of a domain structure because the Kerr microscope images do not reveal any domain formation and the experiment is done under an applied field which should prevail over the possibility of a non-uniform magnetic configuration (see supplementary Figure S5). This decrease is attributed to the reorientation of the magnetization in the plane of the layers. This reorientation seems to be a common feature and supports the hypothesis of the possible precession of the magnetization.

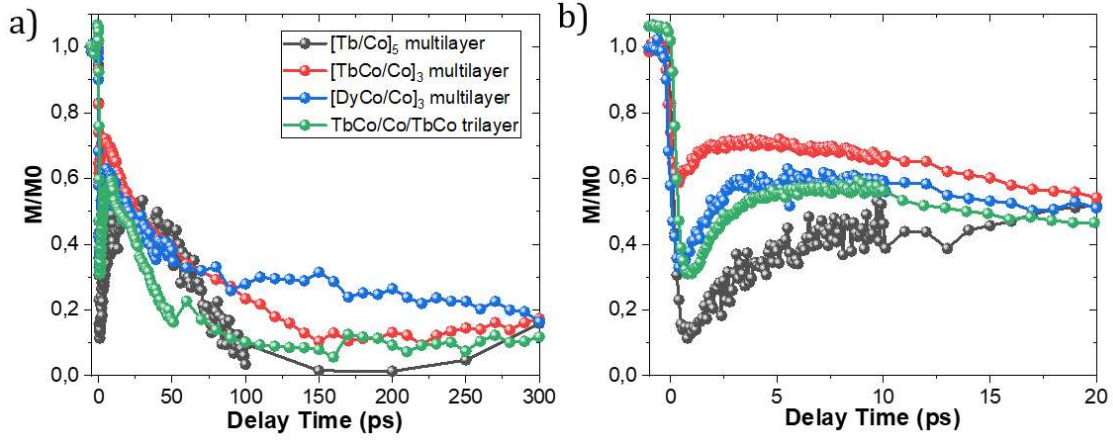

**Figure 4 :** TR-MOKE measurements performed on the different multilayers showing single pulse reversal. a) 300 ps time scale; b) zoom-in of the first 20 ps time scale. The description of the stacks and conditions for TR-MOKE measurements have been show in Table 2.

| Stacks | Composition | Field (Oe) | Energy (µJ) | Repetition rate(kHz) |
|---|---|---|---|---|
| [Tb/Co]$_5$ multilayer | $t$(Tb) = 1.06 nm<br>$t$(Co) = 1.78 nm | 550 | 0.75 | 0.2 |
| [TbCo/Co]$_3$ multilayer | $t$(Tb$_{40}$Co$_{60}$) = 4 nm<br>$t$(Co) = 2.10 nm | 610 | 1.1 | 10 |
| [DyCo/Co]$_3$ multilayer | $t$(Dy$_{30}$Co$_{70}$) = 4 nm<br>$t$(Co) = 2.18 nm | 570 | 1.6 | 1 |
| TbCo/Co/TbCo trilayer | $t$(Tb$_{32}$Co$_{68}$) = 4 nm<br>$t$(Co) = 2.24 nm | 690 | 0.7 | 1 |

**Table2:** Stacks and TR-MOKE conditions corresponding to Figure 4

In this respect, suppressing the source of anisotropy of the Co or weakening its contribution is a key ingredient for robust AOS. This is achieved in the tri-layer in which Co anisotropy is given by the 4nm thick TM-RE alloy at both interfaces. Otherwise, as in a MTJ, if one Co layer has a strong anisotropy, either a multilayer is needed, or the thickness of the TM-RE has to be increased. We tested these two strategies and the results are reported in Figures S6 to S9. We clearly show on the one hand that increasing the number of bilayers in the multilayer, and thus the number of Co layer in contact with an alloy layer, increases the quality of the reversal. On the other hand in a single TbCo/Co bilayer, the switching is favoured when the thickness of TbCo is large compared to the Co layer. This allows obtaining a reversal in a multilayer with a record thickness of 40 nm.

While the in-plane reorientation appears as a key condition for reversal, it is not sufficient to explain a coherent precession. During a reorientation of magnetization, no in-plane direction is a priori favoured which should lead to incoherent rotation and a multidomain state structure after remagnetization in the perpendicular direction. To obtain the observed ring structure, the reorientation must be launched coherently. This could be done by an effective field around which a coherent precession would take place [DAV19, SHE18]. Then, depending on the amplitude of the effective field and the time before the recovery of the effective perpendicular anisotropy, the magnetisation will either end up or down.

Since no field is applied in our single shot experiments, one needs to find the cause of the magnetisation precession and if an effective in plane field could be present in our experiments. One could ascribe this field to shape anisotropy of the thin film, but shape anisotropy can only induce an easy plane of magnetization, not an easy axis in plane. In order for switching to take place, an additional in-plane effective field must exist in the sample, and importantly, this field must remain present when the magnetization is approaching the equatorial plane. In our samples, we first suspected wedges to induce some in-plane anisotropy that could be responsible for this eventual field. Therefore, we fabricated similar films of fixed thickness by rotating the sample during deposition. The ring structures could still be observed after laser exposure. While easy magnetization axes not correlate to deposition conditions in rotation are highly improbable, their existence cannot be excluded at this point.
Another source of symmetry breaking that we can see in the system is the laser's Gaussian profile. Although the laser profile has a cylindrical symmetry, exposure is not homogeneous. Therefore lateral gradients of energy are created in the film. In the static domain structure, we analyse the position of the domain walls that separate the rings. We can observe that the walls are always located at the same local threshold energy and do not depend on their distance to the Gaussian beam's center (see figure S10). The gradients of energy do not seem to dominate the switching mechanism, the locally absorbed fluence does. We can therefore neglect dipolar interactions or lateral induced strain as a possible source for this in-plane effective field. However, this observation is again compatible with a precessional reversal mode.

While the origin of the in plane effective field around which magnetization processes could not be found, how could the coherence of the rotation be explained. The latter is possible if a spin polarized current is established at the first moments of the precession and if it exerts a torque on the magnetization. Indeed, it has been shown that using femtosecond laser-pulse excitation, a spin current is generated in a first ferromagnetic layer of a spin-valve, which can exert a spin-transfer torque on the magnetization of a second spin valve layer, exciting standing spin waves [LAL19]. These spin currents should also exist without Cu, which is the case in our samples, and are known to be dominant during the very short timescales (first ps) [MAL08, IIH08, IGA20, REM20].

In conclusion, single pulse all-optical reversal has been extended to many different materials by the design of optimized heterostructures. The picture of precessional reversal emerges by the strong evidence that magnetization has to reorient in plane before magnetization reverses. The resulting state diagram resembles neither the state diagram observed in the case of AO-HIS in GdFeCo nor the state diagram observed in the case of multiple pulse AO-HDS and demonstrates the existence of a new type of AOS. In terms of reversal time, this precessional like AO-HIS possesses typical time scales between those of GdFeCo and multiple pulse AO-HDS. While the in-plane reorientation of the magnetization has been proven, the origin of the precession must be confirmed and this will be facilitated by the great diversity of the multilayers presenting this effect and whose properties can be adjusted to the experiments to be carried out.

**Acknowledgements**

We thank Bert Koopmans, Jeff Bokor, Martin Weinelt, Daniel Lacour, Nicolas Bergeard, Eric E. Fullerton for fruitful discussions. We also acknowledge financial support from the ANR (ANR-17-CE24-0007 UFO project), the Region Grand Est through its FRCR call (NanoTeraHertz and RaNGE projects), by the impact project LUE-N4S part of the French PIA project "Lorraine Université d'Excellence", reference ANR-15IDEX-04-LUE and by the "FEDER-FSE Lorraine et Massif Vosges 2014-2020", a European Union Program. D.S. has received funding from the European Union's Horizon 2020 research and innovation programme under Marie Skłodowska-Curie grant agreement No 861300 (COMRAD). W. Z gratefully acknowledge the National Natural Science Foundation of China (Grants No. 12104030), China Postdoctoral Science Foundation (Grants No.2022M710320) and China Scholarship Council for their financial support of this work.

**Author contributions**

MH, SM, LP and RS conceived the study. MH and DS fabricated the samples and MH optimized them. YP performed magnetic measurements, YP, DS, JX. and WZ the fast optics measurements with the help of GM, JH, MV and JG. L.D.B-P and DS performed simulations. MH and GM wrote the manuscript with input from all authors. All the authors participated to the scientific discussions.

**Financial Interests.** The authors declare no competing financial interests.

**References**

[KAT00] J. A. Katine, F. J. Albert, R. A. Buhrman, E. B. Myers, and D. C. Ralph, Phys. Rev. Lett. 84, 3149 (2000).
[MIR11] I. M. Miron, K. Garello, G. Gaudin, P.-J. Zermatten, M. V. Costache, S. Auffret, S. Bandiera, B. Rodmacq, A. Schuhl, P. Gambardella, Nature 476, 189–193 (2011).
[LIU10] H. Liu, D. Bedau, D. Backes, J. Katine, J. Langer, A. Kent, Appl. Phys. Lett. 97, 242510 (2010).
[GAR14] K. Garello, C. O. Avci, I. M. Miron, M. Baumgartner, A. Ghosh, S. Auffret, O. Boulle, G. Gaudin, P. Gambardella, Appl. Phys. Let. 105 (2014).
[BIG96] E. Beaurepaire, J.-C. Merle, A. Daunois, J.-Y.Bigot, Phys. Rev. Lett. 76, 4250 (1996).
[STA07] C. D. Stanciu, F. Hansteen, A. V. Kimel, A. Kirilyuk, A. Tsukamoto, A. Itoh, and Th. Rasing, Phys. Rev. Lett. 99, 047601 (2007).
[RAD11] I. Radu, K. Vahaplar, C. Stamm, T. Kachel, N. Pontius, H. A. Durr, T. A. Ostler, J. Barker, R. F. L. Evans, R. W. Chantrell, A. Tsukamoto, A. Itoh, A. Kirilyuk, T., Rasing, A. V. Kimel, Nature 472, 205-209 (2011).
[OST12] T. A. Ostler, J. Barker, R. F. L. Evans, R. W. Chantrell, U. Atxitia, O. Chubykalo-Fesenko, S. El Moussaoui, L. Le Guyader, E. Mengotti, L. J. Heyderman, F. Nolting, A. Tsukamoto, A. Itoh, D. Afanasiev, B. A. Ivanov, A. M. Kalashnikova, K. Vahaplar, J. Mentink, A. Kirilyuk, A. V. Kimel, Nature Communications, 3, 666 (2012).
[STE11] D. Steil, S. Alebrand, A. Hassdenteufel, M. Cinchetti, M Aeschlimann, Phys. Rev. B 84(22), 224408 (2011)
[GOR16] J. Gorchon, R. B. Wilson, Y. Yang, A. Pattabi, J. Y. Chen, L. He, J. P. Wang, M. Li, M., J. Bokor, Phys. Rev. B, 94(18), 19–24 (2016).
[YAN17] Y. Yang, R. B. Wilson, J. Gorchon, C.-H. Lambert, S. Salahuddin, J. Bokor, Science Advances, 3(11), e1603117 (2017).

[DAV20a] C. S. Davies, T. Janssen, J. H. Mentink, A. Tsukamoto, A. V. Kimel, A. F. G. Van Der Meer, A. Stupakiewicz, A. Kirilyuk, Phys. Rev. Appl. 13, 024064 (2020).
[WEI21] W. Zhang, J. X. Lin, T. X., Huang, G. Malinowski, M. Hehn, Y. Xu, S. Mangin, W.. Zhao, Phys. Rev. B 105, 054410 (2022).
[LAL19] M. L. M. Lalieu, R. Lavrijsen, R. A. Duine, and B. Koopmans, Phys. Rev. B 99, 184439 (2019)
[LAL17] M. L. Lalieu, M. J. G. Peeters, S. R. R. Haenen, R. Lavrijsen, B. Koopmans, Phys. Rev. B 96, 220411 (2017).
[BAN20] C. Banerjee, N. Teichert, K. E. Siewierska, Z. Gercsi, G. Y. P. Atcheson, P. Stamenov, K. Rode, J. M. D. Coey & J. Besbas, Nature Communications 11, 4444 (2020)
[ATX14] U. Atxitia, J. Barker, R. W. Chantrell, Phys. Rev. B 89, 224421 (2014).
[DAV20b] C. S. Davies, G. Bonfiglio, K. Rode, J. Besbas, C. Banerjee, P. Stamenov, J. M. D. Coey, A. V. Kimel, A. Kirilyuk, Phys. Rev. Research 2, 032044 (2020).
[AVI19] L. Avilés-Félix, L. Álvaro-Gómez, G. Li, C. S. Davies, A. Olivier, M. Rubio-Roy, S. Auffret, A. Kirilyuk, A. V. Kimel, T. Rasing, L. D. Buda-Prejbeanu, R. C. Sousa, B. Dieny, I. L. Prejbeanu, AIP Advances 9, 125328 (2019).
[AVI20] L. Avilés-Félix, A. Olivier, G. Li, C. S. Davies, L. Álvaro-Gómez, M. Rubio-Roy, S. Auffret, A. Kirilyuk, A. V. Kimel, Th. Rasing, L. D. Buda-Prejbeanu, R. C. Sousa, B. Dieny & I. L. Prejbeanu, Scientific Reports 10, 5211 (2020).
[ALE12] S. Alebrand, M. Gottwald, M. Hehn, D. Steil, M. Cinchetti, D. Lacour, E. E. Fullerton, M. Aeschlimann, and S. Mangin, Appl. Phys. Lett. 101, 162408 (2012); https://doi.org/10.1063/1.4759109
[HAN89] P. Hansen, C. Clausen, G. Much, M. Rosenkranz, K. Witter, J. Appl. Phys. 66, 756 (1989).
[HU22] Z. Hu, J. Besbas, R. Smith, N. Teichert, G. Atcheson, K. Rode, P. Stamenov, J. M. D. Coey, Appl. Phys. Lett. 120, 112401 (2022).
[DAV19] C. S. Davies, K. H. Prabhakara, M. D. Davydova, K. A. Zvezdin, T. B. Shapaeva, S. Wang, A. K. Zvezdin, A. Kirilyuk, Th. Rasing, A. V. Kimel, Phys. Rev. Lett. 122, 027202 (2019).
[SHE18] L. A. Shelukhin, V. V. Pavlov, P. A. Usachev, P. Yu. Shamray, R. V. Pisarev, and A. M. Kalashnikova, Phys. Rev. B 97, 014422 (2018).
[MAL08] G. Malinowski, F. Dalla Longa, J. H. H. Rietjens, P. V. Paluskar, R. Huijink, H. J. M. Swagten, B. Koopmans, Nat. Phys. 4, 855–858 (2008).
[IIH18] S. Iihama, Y. Xu, M. Deb, G. Malinowski, M. Hehn, J. Gorchon, E. E Fullerton, S. Mangin, Adv. Mater. 30, 1804004 (2018).
[IGA20] J. Igarashi, Q. Remy, S. Iihama, G. Malinowski, M. Hehn, J. Gorchon, J. Hohlfeld, S. Fukami, H. Ohno, S. Mangin, Nano Lett. 20, 8654–8660 (2020).
[REM20] Q. Remy, J. Igarashi, S. Iihama, G. Malinowski, M. Hehn, J. Gorchon, J. Hohlfeld, S. Fukami, H. Ohno, S. Mangin, Adv. Sci. 7, 2001996 (2020).
[KIC19] G. Kichin, M. Hehn, J. Gorchon, G. Malinowski, J. Hohlfeld, S. Mangin, Physical Review Applied 12 (2), 024019 (2019).
[WOL09] G. Woltersdorf, M. Kiessling, G. Meyer, J.-U. Thiele, and C. H. Back, Phys. Rev. Lett. 102, 257602 (2009).
[GUA18] G. Lu, X. Huang, S. Fan, W. Ling, M. Liu, J. Li, L. Jin, L. Pan, Journal of Alloys and Compounds 753, 475-482 (2018)
[MOR17] R. Moreno, T. A. Ostler, R. W. Chantrell, O. Chubykalo-Fesenko, Phys. Rev. B 96, 014409 (2017).
[FRI20] B. Frietsch, A. Donges, R. Carley, M. Teichmann, J. Bowlan, K. Döbrich, K. Carva, D. Legut, P. M. Oppeneer, U. Nowak, and M. Weinelt, Sci. Adv. 6, eabb1601 (2020).

[KAT16] V. N. Kats, T. L. Linnik, A. S. Salasyuk, A. W. Rushforth, M. Wang, P. Wadley, A. V. Akimov, S. A. Cavill, V. Holy, A. M. Kalashnikova, and A. V. Scherbakov, Phys. Rev. B 93, 214422 (2016).

# Supplementary figures

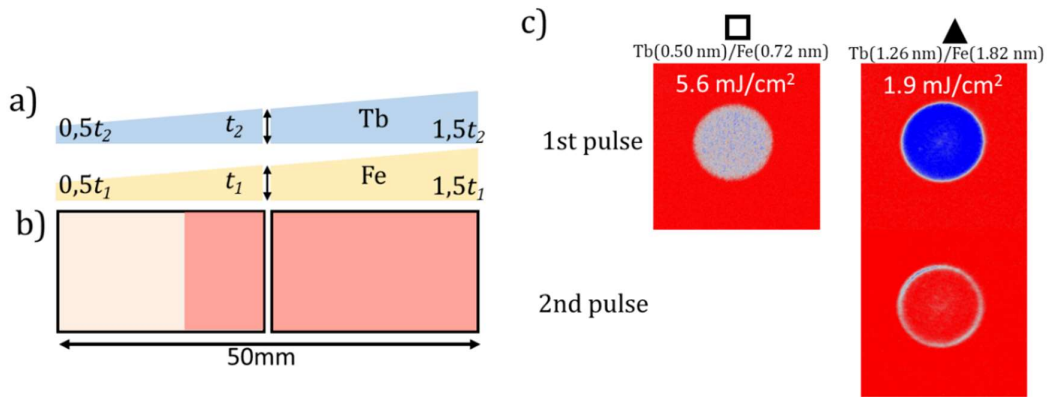

**Figure S1**: Single pulse switching in [Tb(wedge)/Fe(wedge)]$_4$ multilayer. a) Sketch of the multilayer and wedge of Tb and Fe thicknesses, the Fe thickness in the middle is $t_1 = 1.3$ nm, while the Tb thickness in the middle $t_2 = 0.9$ nm. The thickness ratio of Fe and Tb is constant and equals to 1.44 everywhere. b) Light red color shows the region of Tb and Fe thicknesses (1.12 nm < $t_{Fe}$ < 1.96 nm, 0.77nm < $t_{Tb}$ < 1.36 nm) where single shot switching occurs. c) The response to single laser pulse at different positions (pulse duration: 50 fs).

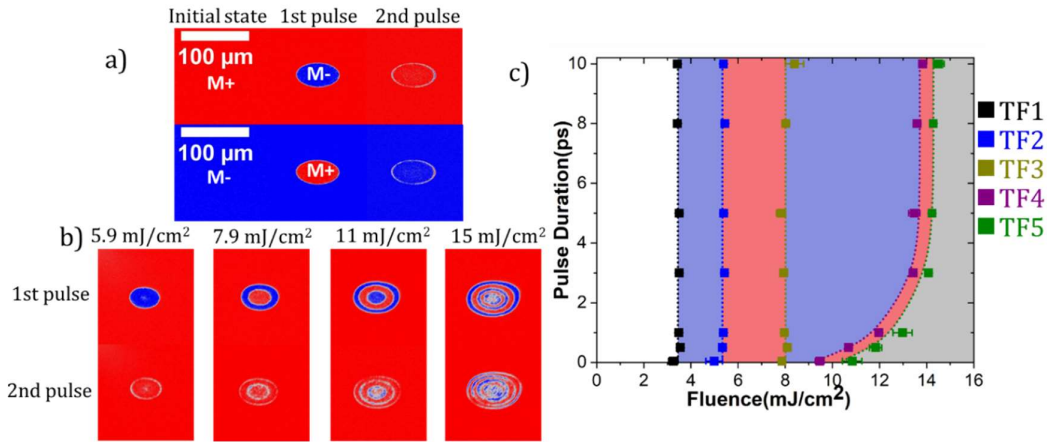

**Figure S2-a:** Single switching and state diagram in [Tb(1.06 nm)/Co(1.78 nm)]$_5$ multilayer. a) Background subtracted images after each single pulse with 5.1 mJ/cm$^2$ laser pulse at 50fs; b) Background subtracted images after first single with 50 fs laser pulse of different fluence; c) State diagram pulse duration versus laser fluence. TF1 is the border for switching; TF2 is the border to two domains state, one ring; TF3 is the border to three domains state, two rings; TF4 is the border to four domains state, three rings; TF5 is the border to demagnetized state at the center and four rings. The dotted lines are guide for the eyes.

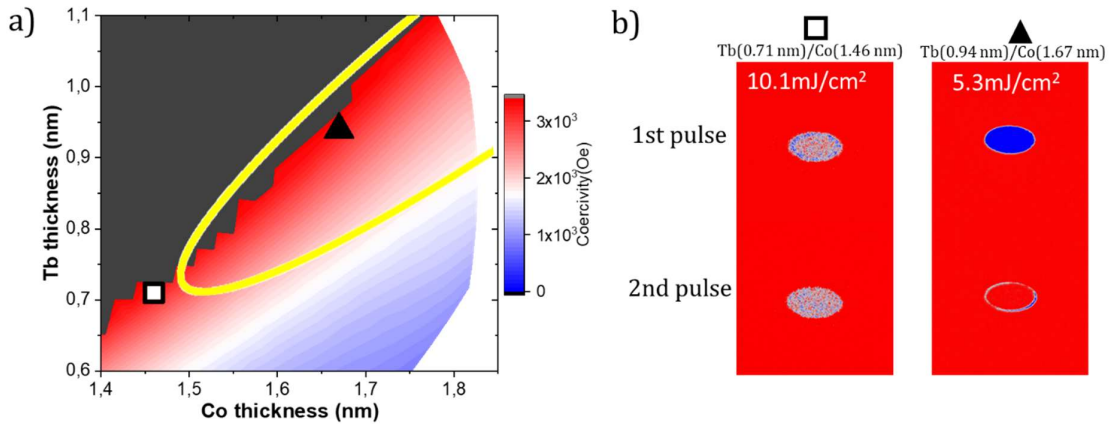

**Figure S2-b:** Single switching in [Tb(wedge)/Co(wedge)]$_5$ cross-wedge multilayer. a) Coercivity map of [Tb/Co]$_5$ along the Tb and Co thickness wedges. The grey color indicates the region in which the applied field was not sufficient to saturate the stack. b) Background subtracted images achieved using 50 fs laser pulse at two positions enclosed by the symbols in a).

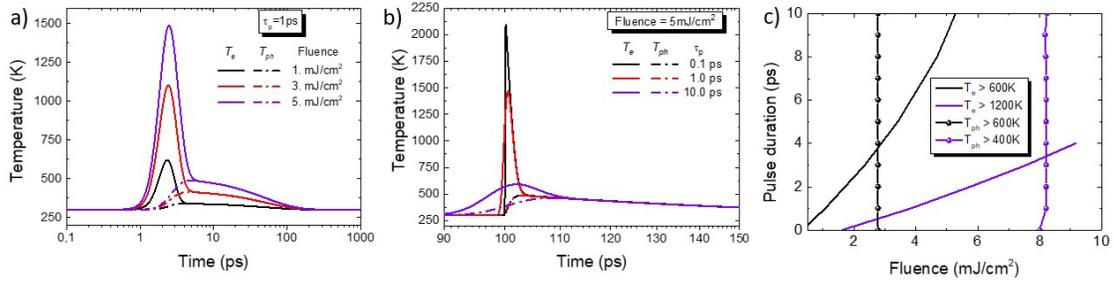

**Figure S3:** Time evolution of the $T_e$ and $T_{ph}$, according with 2T model a) for constant pulse duration of 1ps and variable fluence and b) constant fluence of 5.0 mJ/cm² and variable pulse duration respectively. Simulation parameters: $t_{FM} = 100\ nm$, $T_0 = 300K$, $G = 2.5 \cdot 10^{17}\ W \cdot m^{-3} K^{-1}$, $C_e = \gamma T_e = 225\ Jm^{-3}K^{-2}Te$ and $C_{ph} = 2.6 \cdot 10\ Jm^{-3}K^{-1}$ [MOR17] [FRI20]. c) The fluence threshold $F_{th}$ per pulse duration in order to reach $T_e > 1200K, 600K$ (black and violet lines respectively) and $T_{ph} > 600K, 400K$ (black and violet dotted lines respectively).

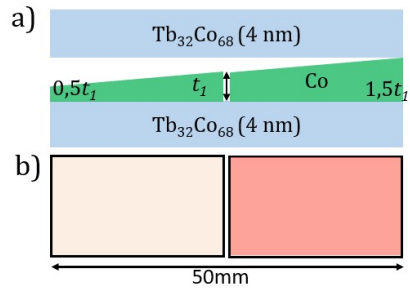

**Figure S4-a:** Single switching in $Tb_{32}Co_{68}$/Co(wedge)/$Tb_{32}Co_{68}$ trilayer. a) Simple stack and wedge of Co thickness description, where the Co thickness in the middle $t_1$ = 1.5 nm. The thickness of Co varies from 0.83 nm (left) to 2.27 nm (right); b) Light red color shows the region of Co thickness (1.52 nm < $t_{Co}$ < 2.27 nm) where single shot switching occurs in $Tb_{32}Co_{68}$/Co/ $Tb_{32}Co_{68}$ trilayer.

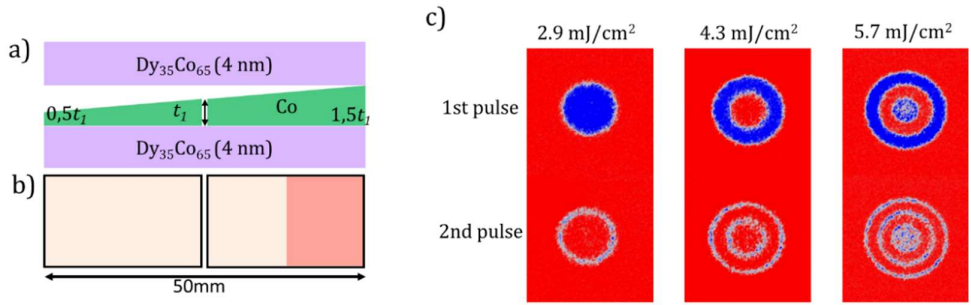

**Figure S4-b:** Single switching in $Dy_{35}Co_{65}$/Co(wedge)/$Dy_{35}Co_{65}$ trilayer. a): Simple stack and wedge of Co thickness description, where the Co thickness in the middle $t_1$ = 2 nm. The thickness of Co varies from 1.02 nm (left) to 3.02 nm (right); b) Light red color shows the region of Co thickness (2.41 nm < $t_{Co}$ < 3.02 nm) where single shot switching occurs in $Dy_{35}Co_{65}$/Co/$Dy_{35}Co_{65}$ trilayer; c) Background subtracted images after each 50fs laser pulse with different fluences, where $t_{Co}$ = 2.98 nm.

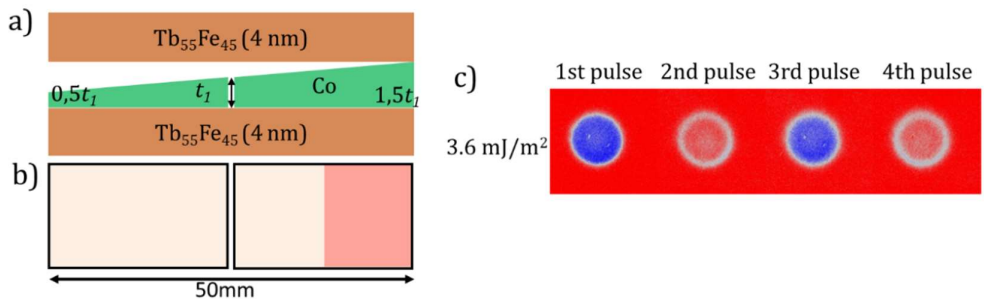

**Figure S4-c**: Single switching in $Tb_{55}Fe_{45}$(4 nm)/Co(wedge)/$Tb_{55}Fe_{45}$ (4 nm) trilayer by 50 fs single pulse. a) Simple stack and wedge of Co thickness description, where the Co thickness in the middle $t_1$ = 1.5 nm. The thickness of Co varies from 0.83 nm (left) to 2.27 nm (right); b) Light red color shows the region of Co thickness (3.89 nm < $t_{Co}$ < 5.60 nm) where single pulse switching occurs. c) Background subtracted images after each 50fs laser pulse and fluence of 3.6 mJ/cm² measured at the green point position with $t_{Co}$ = 2.27 nm
1717

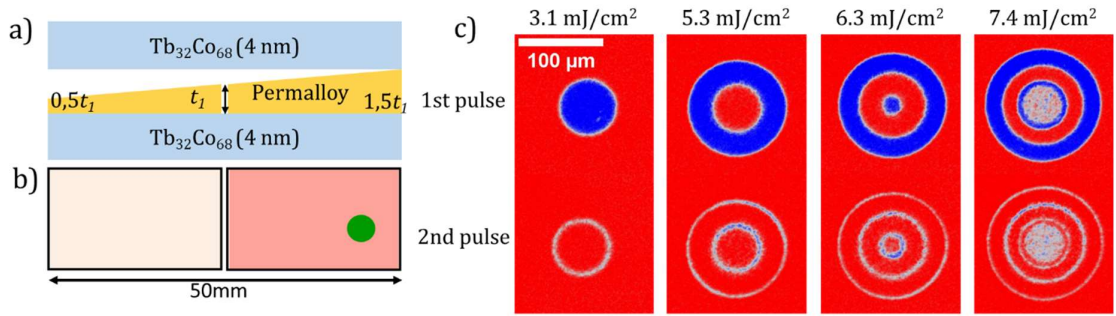

**Figure S4-d:** Single switching in $Tb_{32}Co_{68}$/Permalloy(wedge)/$Tb_{32}Co_{68}$ trilayer. a): Simple stack and wedge of Py thickness description, where the Py thickness in the middle $t_1 = 3.7$ nm. The thickness of Py varies from 2.05 nm (left) to 5.60 nm (right); b) Light red color shows the region of Py thickness (3.89 nm < $t_{Py}$ < 5.60 nm) where single shot switching occurs; c) Background subtracted images after each 50fs laser pulse with different fluences measured at the green point position with $t_{Py}$ = 5.10 nm.

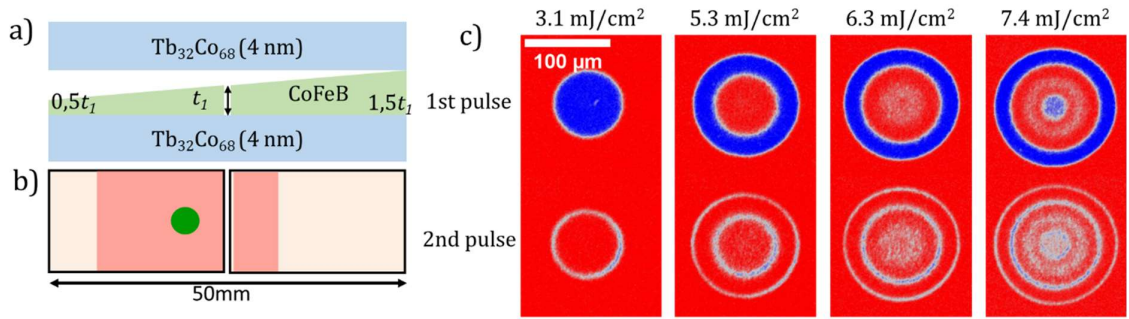

**Figure S4-e:** Single switching in $Tb_{32}Co_{68}$/CoFeB(wedge)/$Tb_{32}Co_{68}$ trilayer. a): Simple stack and wedge of CoFeB thickness description, where the CoFeB thickness in the middle $t_1 = 2.5$ nm. The thickness of CoFeB varies from 1.39 nm (left) to 3.78 nm (right); b) Light red color shows the region of CoFeB thickness (1.67 nm < $t_{CoFeB}$ < 2.87 nm) where single shot switching occurs; c) Background subtracted images after each 50fs laser pulse with different fluences measured at the green point position with $t_{CoFeB}$ = 3.33 nm.

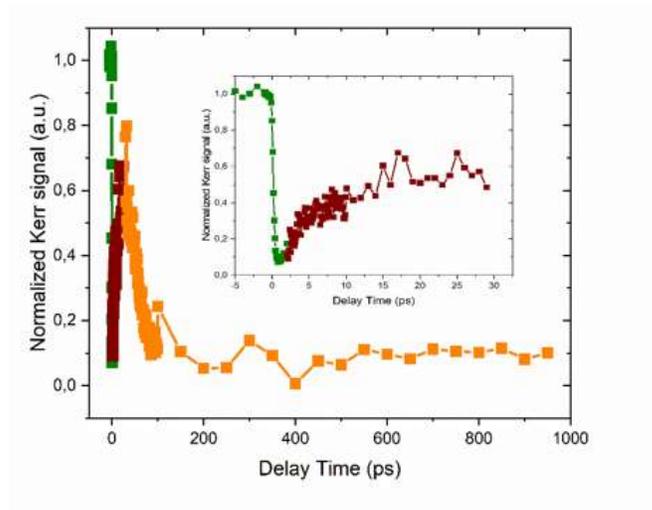
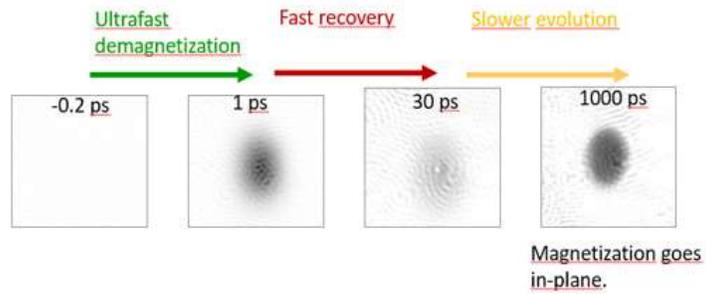

**Figure S5:** Demagnetization dynamics in [Tb(1.06 nm)/Co(1.78 nm)]$_5$ multilayer. The Kerr images show that magnetization decrease is not due to the appearance of a domain pattern.

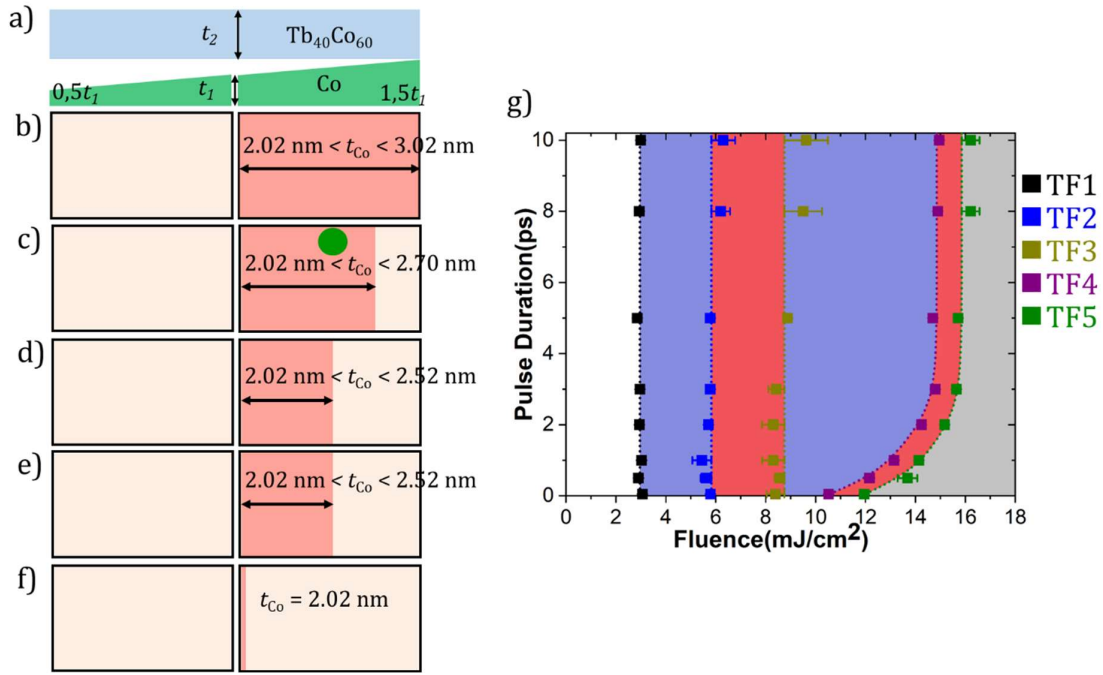

**Figure S6**: Single pulse switching in [Co$_{60}$Tb$_{40}$)(x)/Co(wedge)]$_3$, x varying from 1 to 5nm; Thickness mapping of Co layer that show single switching in [Tb$_{40}$Co$_{60}$/Co]$_3$ multilayer. a) Description of Co wedge. The thickness of Co layer is a wedge where the thickness in the middle $t_1$ = 2 nm; (b-e) Light red color shows the region where single switching occurs when the thickness of TbCo layer is b) $t_2$ = 5 nm; c) $t_2$ = 4 nm; d) $t_2$ = 3 nm; e) $t_2$ = 2 nm; f) $t_2$ = 1 nm. g) State diagram obtained at green round position in c), where $t_{Co}$ = 2.52 nm.

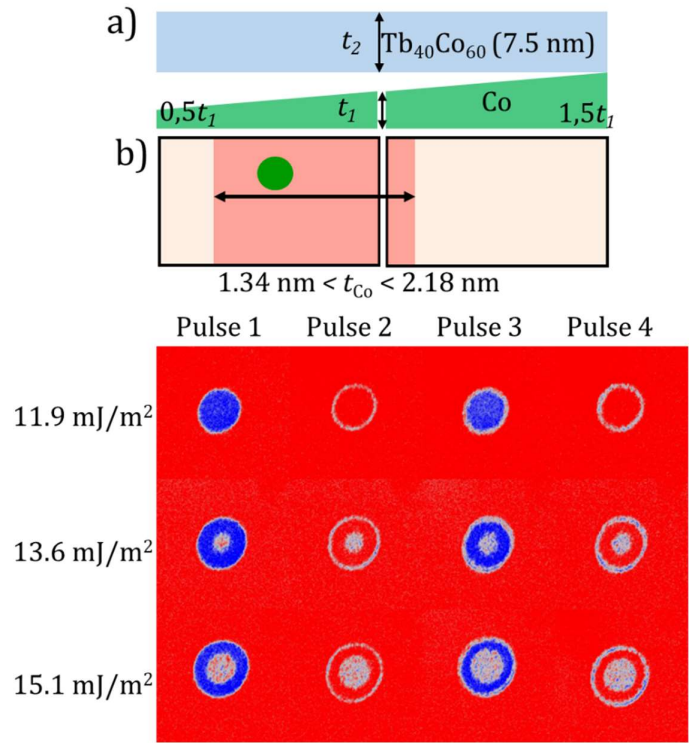

**Figure S7 :** Single pulse switching in [Tb$_{40}$Co$_{60}$)(7.5 nm)/Co(wedge)]$_3$ multilayer; a): Simple stack and wedge of Co thickness description, where the Co thickness in the middle $t_1$ = 2 nm. The thickness of Co varies from 1.11 nm (left) to 3.02 nm (right); b) Light red color shows the region of Co thickness (1.34 nm < $t_{Co}$ < 2.18 nm) where single shot switching occurs; c) Background subtracted images after each 50fs laser pulse with different fluences obtained at green round position in b), where $t_{Co}$ = 1.57 nm.

___

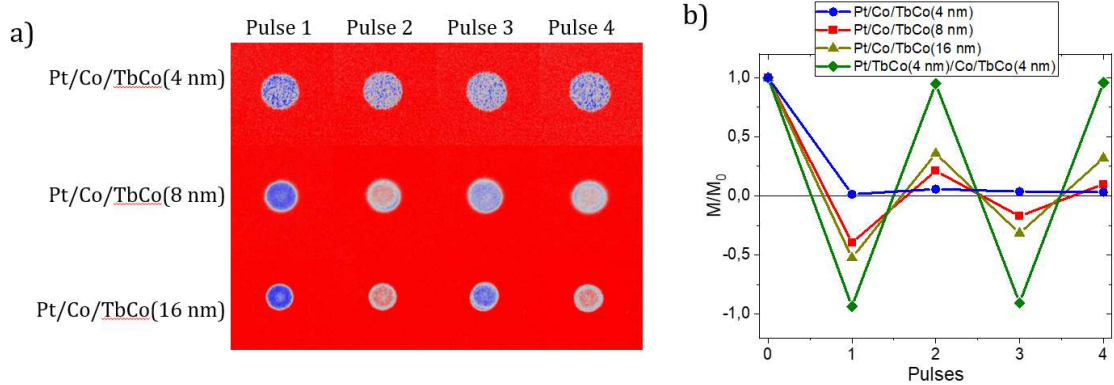

**Figure S8 :** a): Single pulse switching in Co(2.27 nm)/Tb$_{32}$Co$_{68}$ (x) bilayer where x is 4 nm (top), 8 nm (middle) and 16 nm (bottom). b) The changes of magnetization around the center of the spot, estimated from magnetic contrast, are plotted as a function of number of pulses. The value of Tb$_{32}$Co$_{68}$ (4 nm) /Co(2.27 nm)/Tb$_{32}$Co$_{68}$ (4 nm) is given for comparison.

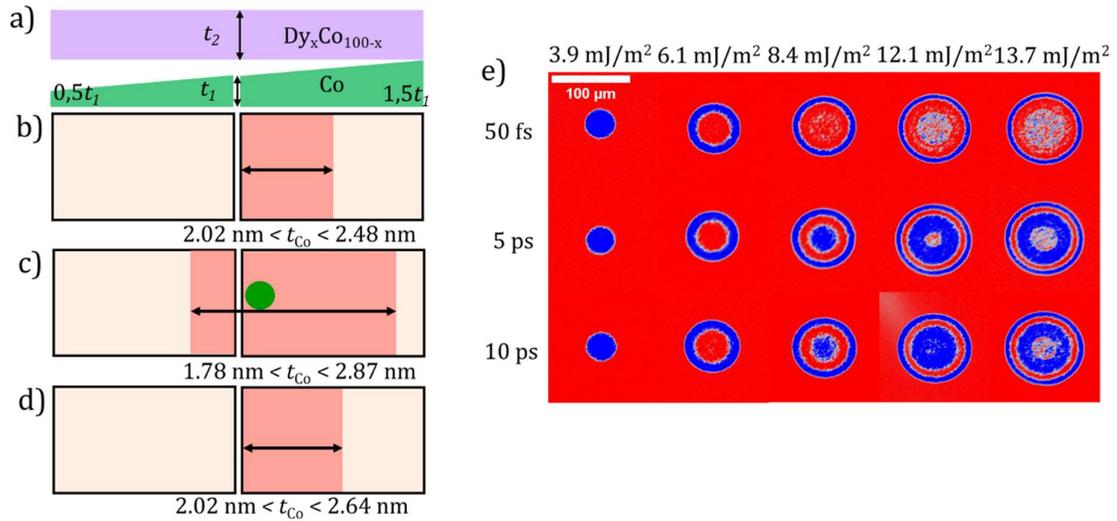

**Figure S9 :** Single pulse reversal in [Dy$_{1-y}$Co$_y$(4 nm)/Co(wedge)]$_3$ for y = 65, 70 and 75%. Thickness mapping of Co layer that show single switching. a) Description of Co wedge. The thickness of Co layer is a wedge where the thickness in the middle $t_1$ = 2 nm. The thickness of DyCo alloy layer is kept constant at $t_2$ = 4 nm. Light red color shows the region of Co thickness where single shot switching occurs in b) Dy$_{35}$Co$_{65}$; c) Dy$_{30}$Co$_{70}$; d) Dy$_{25}$Co$_{75}$. e) State diagram obtained at green round position in c), where $t_{Co}$ = 2.10 nm.

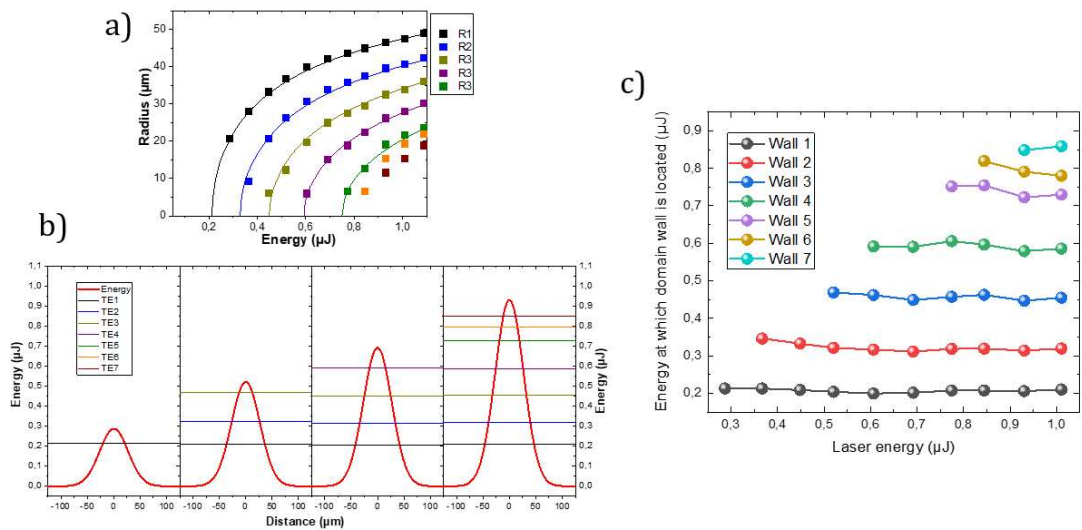

**Figure S10 :** Analysis of the domain wall position of the ring structure, in space and energy, versus laser pulse energy in $Tb_{32}Co_{67}$(4 nm)/Co(1.52 nm)/$Tb_{32}Co_{67}$(4 nm) trilayer. a) Ring radius versus energy. b) Plot of the gaussian laser profile for different laser energies and report of the ring radius. c) position in energy of the domain walls versus total laser energy.

# Supplementary materials

**Deposition conditions and wedge fabrication**

The multilayers were produced by sputtering in an AJA machine. The alloys are made by co-deposition of different sources arranged in a confocal geometry, the composition being adjusted by the powers applied on the different cathodes. The homogeneity of the layer is obtained by rotating the sample.

Thickness wedges are obtained by stopping the rotation, resulting in a gradient of thickness. A calibration of the thicknesses as a function of the position has been done beforehand, which allows to study the magnetic properties as a function of the thickness in a controlled way and on a single sample.

**Laser set up and fluence measurements**

**AOS Measurement.** Ti:sapphire femtosecond-laser source and regenerative amplifier were used for the pump laser beam in AOS measurement. Wavelength and repetition rate of the femtosecond laser are 800 nm and 5 kHz, respectively. LED light probe source with a wavelength of 628 nm was used for taking MOKE images. The Gaussian beam size was determined in two ways: by directly observing the beam at the focal plane of the microscope's lens (Used for imaging) and by using the domain size vs pulse energy fit (See data analysis). For samples grown on glass substrate, pump laser excitation was done on one side of the sample and MOKE microspore observation was on the other side (Transmission configuration). Both methods are possible to determine the beam side and provide consistent results. For samples grown on silicon substrate, the pump and probe light are both on the sample side (Reflection configuration). In this case, the beam size was only obtained by fitting the domain size vs pulse energy.

**Data analysis**. The measured laser incident fluence was calculated using

$$\bar{F} = \frac{P}{fS} \quad (S1)$$

where $P$ is the measured power, $f$ is the repetition rate of the laser ($f$ = 5 kHz), and $S$ is the beam spot area.

For transmission configuration, the beam spot area is

$$S = \pi R^2 \quad (S2)$$

where R is the radius of the beam spot and equals to $\frac{\sqrt{2}}{2}$FWHM, where FWHM is half of the laser spatial full-width half-maximum. So, the measured fluence $\bar{F}$ is defined such that the power of the laser is divided by a factor $e^2$ at a distance from the center of the beam:

$$\bar{F} = \frac{P}{f\pi R^2} \quad (S3)$$

The fluence profile has then the form:

$$F(r) = 2\bar{F}e^{-(\frac{r}{R})^2} \quad \text{(S4)}$$

where $r$ is the domain radius. The peak value is then twice its measured value. A magnetic domain starts to appear as soon as the fluence peak value exceeds a certain threshold value $F_{th}$. When this happens, one would measure a fluence $\bar{F} = \frac{F_{th}}{2}$. The domain radius must the verify the following equation:

$$F(r) = F_{th} \leftrightarrow r = R\sqrt{\frac{1}{2}\ln(\frac{\bar{F}}{\bar{F}_{th}})} \quad \text{(S5)}$$

Considering equation S1, the threshold powers as well as beam spot radius can be extracted as a fitting parameter:

$$r = R\sqrt{\frac{1}{2}\ln(\frac{P}{P_{th}})} \quad \text{(S6)}$$

For the reflection configuration, the domains as well as the beam spot are elongated due to an angle of incidence of 45°. We used the following equation to extract the threshold powers and beam spot area as a fitting parameter: Domain area = $S\ln(\frac{P}{P_{th}})$

**Modelling of electron and phonon temperature to explain the specific state diagram**

The specific features of the shape of the experimental state diagrams can be understood in the frame of a two-temperature (2T) model [MOR17]. If we assume that all optical switching is due to a rapid increase of temperature above a certain threshold, we can draw some interesting points.

The standard 2T model uses the electron and phonon bath to describe the heat components in the system. Each bath has its own heat capacity $C_i (i = e, ph)$ and exchanges energy through the coupling parameter $G$. $P(t)$ describes the laser pulse excitation that heats the electron bath and the two coupled equations read as:

$$C_e \frac{dT_e(t)}{dt} = -G[T_{ph}(t) - T_e(t)] + P(t) - \frac{C_e}{\tau_0}[T_e(t) - T_0]$$

$$C_p \frac{dT_{ph}(t)}{dt} = -G[T_e(t) - T_{ph}(t)] - \frac{C_{ph}}{\tau_0}[T_{ph}(t) - T_0]$$

$$P(t) = \frac{F}{t_{FM}\tau_L} \cdot e^{\left[-\frac{(t-t_0)^2}{\tau_L^2/(4ln2)}\right]}$$

Here $\tau_L$ is the laser pulse duration, F is its fluence, $t_0$ is the moment of the laser pulse application, $t_{FM}$ is the thickness of the magnetic film, $\tau_0$ is the cooling characteristic time pulse to recover the initial temperature of the system $T_0$.

In Fig.S3 shows the time evolution of the electron temperature $T_e$ and phonon temperature $T_{ph}$ respectively upon varying the laser fluence and/or pulse duration. As expected, increasing fluence rises both electron and phonon temperatures but, interestingly, with constant fluence the electron temperature peak reduces drastically upon increasing the pulse duration (e.g. from 100fs to 1ps at $F$ = 5.0 mJ/cm² the $T_e$ peak drops of ~500 K).

The AO-HIS mechanism in GdFeCo alloy has been related to a difference in the characteristic demagnetization time of the two sub-lattices (Gd and Fe/Co) that have to demagnetise almost completely in order to switch. Because the TM sub-lattice has the higher $T_c$ and its magnetic properties come from the 3d electrons, it makes sense that, to switch, the electron temperature $T_e$ needs to approach $T_c$ and thus due to dependency on the pulse duration shown in Fig.1, the resulting state diagram has a triangular shape [WEI21]. In contrast, for a reorientation of magnetization and precessional switching it is not required to completely demagnetize the TM, or at least it is not crucial, but what seems to be necessary is that the phonon temperature $T_{ph}$ reaches a certain limit value. The dependency on the phonon temperature instead of the electron temperature $T_e$ is justified by the strong spin orbit coupling of the Tb, contrary from Gd. Fig. S3c shows the threshold fluence needed to reach a limit value either for the electron temperature $T_e$ or the phonon temperature $T_{ph}$. The difference between the two trends is similar with our experimental state diagrams in [Tb/Co] and GdFeCo [WEI21].